\documentclass[a4]{epl2}
\usepackage{graphicx}
\usepackage{epsfig}
\usepackage{color}
\usepackage{amssymb}

\title{\textcolor[rgb]{0.00,0.00,1.00}{{On Abraham-Lorentz force, Unruh and Hawking radiations}}}
\shorttitle{\textcolor[rgb]{0.00,0.00,1.00}{On Abraham-Lorentz force and Unruh radiations}}

\author{\textcolor[rgb]{0.00,0.00,0.00}{A. I. Arbab}\inst{}\footnote{\textcolor[rgb]{0.00,0.00,0.00}{arbab.ibrahim@gmail.com}}}

\institute{\inst{}
Department of Physics,
College  of Science, Qassim University, P.O. Box 6644, Buraidah 51452, KSA}

\abstract{ Assuming the radiation emitted by an accelerating charge follows the Unruh radiation, we obtained the  characteristics of the de Broglie wave associated with the accelerating charge.  The de Broglie wavelength of the accelerating charged particle is found to be inversely proportional to the temperature of the emitted radiation.  Merging the Abraham-Lorentz and Unruh formulae shows that the particle de Broglie wavelength is found to vary inversely with its acceleration. It is found to have the same structure as that of the Wien's displacement law relating the maximum wavelength of the Black Body radiation to its temperature. A maximum acceleration that a charged particle can attain is derived that sets a limit to the the maximum electric field. The Abraham-Lorentz force for a black hole radiation is found to be proportional to its evaporation rate. The final mass of the black hole left-over is found to be $\sqrt{\frac{\alpha\hbar c}{24 \pi G} }$, where $G$ is the gravitational constant, $c$ the speed of light, $h=2\pi\hbar$ is the Planck constant, and $\alpha$ is the fine structure constant. The minimum entropy and spin of the black hole emitting Hawking radiation are, respectively, found to be $(\alpha/6) k_B$ and $(\alpha/6)\hbar$. The presently observed universal acceleration is a manifestation of the  Unruh black body temperature of $10^{-29}K (10^{-33}eV)$. This agrees with the black body radiation temperature ($T$) relating $TR=const.$ prevailing since the time of the big bang, where $R$ is the universe radius.}

\pacs{}{}

\begin{document}
\maketitle
\baselineskip=19pt

\section{\textcolor[rgb]{0.00,0.07,1.00}{Introduction}}

The Newton's laws of motion are found to be fully describe the motion of particles describing the motion of  particles. The second law of motion relates the particle acceleration to the causing force. Most of physical systems undergo a constant acceleration motion. However, when the acceleration is not constant, a jerky motion results in. The Jerk defines the rate of change of the acceleration. The acceleration can  be considered to be a vector field that depends on space and time, or generally on coordinates. The Jerk can be defined in a covariant form, and  its form in curved space can then be deduced. Thus, under general coordinate transformation one can think of a more general definition of the Jerk.

It is shown by Abraham, Lorentz and Dirac that an accelerating charged particle undergoes a reaction force \textcolor[rgb]{0.00,0.07,1.00}{\cite{abraham, lorentz, dirac}}. This force is connected with some characteristic time (of the order of $10^{-24}s$ for an electron). This time can be connected with the jerky  motion the accelerating charge experienced after radiating away the electromagnetic radiation. But it is remarkable that this time is reachable for a gravitating accelerating mass ($10^{-11}s$ for the Earth). At the same time Schrodinger had noticed that the moving electron exhibits a jittery (bewungung) motion with a frequency of $2mc^2/\hbar$, and an amplitude of the order of Compton wavelength, $\hbar/mc$, where $m$ is the mass of the electron, and $h=2\pi\hbar$ is the Planck's constant \textcolor[rgb]{0.00,0.07,1.00}{\cite{schrod}}. This rapid oscillation of the electron is attributed to the interference between the positive and negative energy states of the electron.

An intimate quantum phenomenon of interest is the Unruh effect, where a detector undergoing uniform acceleration $a$ in a vacuum field responds as though it were
immersed in thermal radiation of temperature $T$ given by $T=\hbar a/2\pi k_Bc$, where $k_B$ is the Boltzman constant \textcolor[rgb]{0.00,0.07,1.00}{\cite{unruh}}. This latter effect can be seen as  some kind of the radiation reaction stated above.  Employing quantum mechanics,  Stephen Hawking derived formula stating that black holes also radiate like a black body \textcolor[rgb]{0.00,0.07,1.00}{\cite{hawking}}.  This formula is analogous to the Unruh radiation. As the black hole radiates, its mass decreases and so may eventually evaporate or end with some final critical mass. We thus assume that the Unruh radiation due to accelerating observer (charge) causes a reaction force on the observer as that of the Abraham-Lorentz force. The two effects are apparently due to some misconception of the nature of the electron.  In Dirac quantum theory of the electron, the electron apparently moves with speed of light in two directions. Some authors consider this to usher in the fact that the electron has some internal structure and not very fundamental. Since a moving charged particle has a wave nature, the de Broglie wave, one would expect the quantum aspect to play a significant role too. Bekenstein derived a formula relating the black hole entropy to its horizon area \textcolor[rgb]{0.00,0.07,1.00}{\cite{entropy}}. Since the horizon area, the entropy of a black hole is directly proportional to its mass squared. The entropy is  inversely related to the information contained inside the event horizon of a black hole. A censorship theorem developed by Penrose that forbids the information to be lost inside a black hole \textcolor[rgb]{0.00,0.07,1.00}{\cite{censor}}. Penrose also hypothesized that no naked singularity can exist in our universe.

The presently observed  universal acceleration, that is attributed to  the expansion of space, would produce a thermal Unruh emission. At the same time the black body radiation in the early universe that follows the relation $TR=const.$, where $R$ is the radius of the universe and $T$ is the temperature of the radiation, could still be satisfied for such a radiation. The present universal acceleration is sometimes referred to the Hubble acceleration. Hence, combining the two equations, the present temperature (acceleration)  resulting from the space acceleration (temperature) can be calculated. It would remarkable to attribute the energy connected with this radiation to the cosmological constant (vacuum energy). Therefore, the Unruh radiation will be instrumental to account for the vacuum energy.

We explore in this paper the particle Jerk in curved space. We do so by allowing the acceleration to be a vector field. A covariant form can then be investigated and its form in curved space can be derived. We also connect the reactive force to the jittery motion executed by a traveling  electron. Equating the matter wave frequency to the jitter frequency exhibited by a moving electron,   that was shown by Schrodinger,  we obtained a maximal acceleration that a moving charged can't exceed. Substituting this acceleration in the Larmor formula that gives  the power of electromagnetic wave due to an  accelerating charge, a maximal power is obtained. Similarly,  a maximal  electric field is found that exceeds the Schwinger limit.

\section{\textcolor[rgb]{0.00,0.07,1.00}{The relativistic Jerk}}

The Jerk defines a kinematical quantity that defines the rate of change of acceleration. In nature, most systems are described by constant acceleration. Of such motions are the  motion of a particle under gravity and the motion of a particle executing simple  harmonic motion.

The acceleration in one dimension can be  defined as
\begin{equation}
a_x=\frac{dv_x}{dt}=\frac{dv_x}{dx}\,\frac{dx}{dt}=\frac{1}{2}\,\frac{dv_x^2}{dx}\,.
 \end{equation}
Therefore, the Jerk can then be defined as
\begin{equation}
J_x=\frac{da_x}{dt}=\frac{da_x}{dx}\,\frac{dx}{dt}=v_x\frac{da_x}{dx}\,.
\end{equation}
Upon using Eq.(1), Eq.(2) yields
\begin{equation}
J_x=\frac{1}{2}\,v_x\frac{d^2v^2_x}{dx^2}\,.
\end{equation}
However, if $\vec{a}$ is now treated as a vector field, \emph{i.e.}, a function of $x, y, z, t$, then one has
\begin{equation}
\vec{J}=\frac{d\vec{a}}{dt}=\frac{\partial\vec{a}}{\partial t}+(\vec{v}\cdot\vec{\nabla})\vec{a}\,.
\end{equation}
The presence of a Jerk is generally connected with dissipation in the system.

In a covariant (relativistic) form, one has
\begin{equation}
J^{\mu}=V^\nu\partial_\nu  a^\mu\,\qquad \partial\mu=\left(\frac{1}{c}\frac{\partial}{\partial t}\,,\vec{\nabla}\right)\,,\qquad V^\mu=(\gamma\, c, \gamma\, \vec{v})\,,\qquad \gamma=(1-v^2/c^2)^{-1/2}\,.
\end{equation}
Therefore, one defines the proper Jerk as
\begin{equation}
\frac{Da^\mu}{d\tau}=V^\nu\partial_\nu a^\mu\,.
\end{equation}
Notice that in special relativity, $a^\mu\,V_\mu=0$ implying that in 4-dimensions the acceleration, and velocity are perpendicular to each other (like the circular motion).
Owing to Eq.(5), Eq.(4) is the non-relativistic analogue of Eq.(5). Now one has, $J^\mu=\frac{da^\mu}{d\tau}$, where $dt=\gamma\, d\tau$.

In a curved space, one can write Eq.(5) as
\begin{equation}
J^{\mu}=V^\nu\nabla_\nu  a^\mu\, \,,
\end{equation}
where $\nabla_\nu$ is the covariant derivative, $V^\nu$ is the tangent vector along the geodesic parameterized by $\tau$, \emph{i.e.}, $x^\nu=x^\nu(\tau)$, \emph{viz}., $V^\nu=\frac{dx^\nu}{d\tau}=\dot x^\nu$\,, and that
\begin{equation}
a^\mu=\frac{dV^\mu}{d\tau}\,.
\end{equation}
For null geodesic, one chooses a different parameter than $\tau$. The 4-vector Jerk generalizes the ordinary 3-vector Jerk. The zero-component of the Jerk defines the scalar part of the 4-vector Jerk.  The zero-component of the 4-force  is proportional to the power. However, the physical significance of the zero-component of the Jerk is not yet defined.

The geodesic equation (the straight line analogue of flat space) is defined by
\begin{equation}
\frac{d^2x^\mu}{d\tau^2}+\Gamma^\mu_{\nu\lambda}\dot x^\nu \dot x^\lambda=0\,,\qquad \qquad \frac{dV^\mu}{d\tau}+\Gamma^\mu_{\nu\lambda}V^\nu V^\lambda=0 \,.
\end{equation}
The covariant derivative of the acceleration 4-vector in curved space is defined as
\begin{equation}
\nabla_\nu  a^\mu=\partial_\nu a^\mu+\Gamma^\mu_{\nu\lambda}a^\lambda\, \,.
\end{equation}
Applying Eq.(10) in Eq.(7) yields
\begin{equation}
J^{\mu}=V^\nu\nabla_\nu  a^\mu=V^\nu\partial_\nu a^\mu+\Gamma^\mu_{\nu\lambda}a^\lambda V^\nu\, \,.
\end{equation}
Differentiating Eq.(9) with respect to $\tau$ and using Eq.(8) yield
\begin{equation}
\frac{da^\mu}{d\tau}+K^\mu_{\,\,\,\,\alpha\beta\lambda}\, V^\alpha\,V^\beta\, V^\lambda=0\,,
\end{equation}
where
\begin{equation}
K^\mu_{\,\,\,\,\alpha\beta\lambda}=\partial_\alpha\Gamma^\mu_{\beta\lambda}-2\Gamma^\mu_{\lambda\nu}\,\Gamma^\nu_{\alpha\beta}\,.
\end{equation}
In flat space, $K^\mu_{\,\,\,\,\alpha\beta\nu}$ vanishes and the Jerk reduces to the ordinary case, $\vec{J}=\frac{d\vec{a}}{dt}$.

Equation (11) now reads
\begin{equation}
J^\mu=\frac{da^\mu}{d\tau}+\Gamma ^\mu_{\,\nu\lambda}\, a^\lambda\,V^\nu\,.
\end{equation}
It is interesting that Eq.(14) shows that the Jerk in curved space is proportional to the acceleration too.
Equations (12) and (14) show that in the absence of a Jerk, one has
\begin{equation}
\Gamma ^\mu_{\,\nu\lambda}\, a^\lambda=K^\mu_{\,\,\,\,\alpha\beta\nu}\, V^\alpha\,V^\beta\,.
\end{equation}
Equation (15) can be written as
\begin{equation}
\Gamma ^\mu_{\,\nu\lambda}\, \frac{dV^\lambda}{d\tau}=K^\mu_{\,\,\,\,\alpha\beta\nu}\, V^\alpha\,V^\beta\,,
\end{equation}
which is a nonlinear equation that can be solved to find the velocity, $V^\mu$\, if a metric is given.

\section{\textcolor[rgb]{0.00,0.07,1.00}{The radiation reaction: Abraham - Lorentz force in curved space}}

The Abraham - Lorentz force  is the recoil force on an accelerating charged particle caused by the particle an emitting electromagnetic radiation. It is also called the radiation reaction force. This force is proportional to the charge acceleration. It is defined by \textcolor[rgb]{0.00,0.07,1.00}{\cite{abraham}}
\begin{equation}
\vec{F}=\frac{\mu_0 q^2}{6\pi c}\, \vec{\dot a}\,,
\end{equation}
where $\mu_0$ is the permeability of the free space, and $q$ is the charge of the accelerating particle. This force is some times referred to as a self-force. It is interesting that the this force is independent of the sign of the  charge of the accelerating particle.

In curved space, Abraham-Lorentz equation will read
\begin{equation}
F^\mu=\frac{\mu_0 q^2}{6\pi c}\left[\frac{da^\mu}{d\tau}+\Gamma ^\mu_{\,\nu\lambda}\, a^\lambda\,V^\nu\right]\,,
\end{equation}
upon using Eq.(14). However, in curved space one would expect a 3-dimensional (vector) force to be given by
\begin{equation}
f^i=\frac{\mu_0 q^2}{6\pi c}\, J^i\,,
\end{equation}
where $J^i$ is defined by Eq.(14), and $i=1, 2, 3$, representing the x, y, and z-directions. The physical meaning of temporal component of the Jerk is not yet defined. The  force in Eq.(17) can be expressed as
\begin{equation}
\vec{F}=m\tau_q\, \vec{\dot a}\,,\qquad\qquad \tau_q=\frac{\mu_0 q^2}{6\pi mc}\,,
\end{equation}
  where $\tau_q$ is some characteristic frequency connected with the jolted  motion resulted from emitted electromagnetic radiation.

If we now make a Taylor series expansion for the  acceleration, one finds
$$ a(t)=a_0+\tau_q\dot a+\frac{\tau_q^2}{2}\,\ddot a$$
so that Eq.(20) becomes
$$\vec{F}=m\,\tau_q\dot a+m\frac{\tau_q^2}{2}\,\ddot a $$
For an accelerating electron/positron, one has a very high frequency, $\tau_q^{-1} $, of a value of $1.6\times 10^{23}Hz$ that is associated with this acceleration rate.  A uniformly moving electron is found by Schrodinger to experience a jittery motion with a frequency defined by  $f_0=2mc^2/\hbar=1.55\times 10^{22}Hz$ \textcolor[rgb]{0.00,0.07,1.00}{\cite{schrod}}. Therefore, these two times are roughly equal! However, while the former time is associated with the electron in classical aspect of Maxwell's equations, the latter time is connected with the electron in the  Dirac quantum mechanics. Consequently, one may conclude that the jittery motion is somehow connected with the reaction force that an accelerating charged  particle experiences. Thus, the reason behind such a motion could be due to existence of some vacuum interaction resulting from a violent vibration of the accelerating charge to slow it down.

Owing to  quantum mechanics, a moving electron exhibits a matter wave nature. On the other hand, the nature of the matter wave connected with an accelerating mass is not known. If such a radiation exists, then the mechanics by which an accelerating mass emits matter radiation is not yet addressed. When the mass moves, it experiences an inertia. The legitimate question is that does the matter wave exhibits an inertia too? Does the matter radiation has the Black Body characteristics? We raise the question that is the particle matter wavelength (frequency) resonate with the electromagnetic wave emitted by the charge? Moreover, how the de Broglie wavelength is related to the acceleration of the particle?

A relativistic generalization of the Abraham-Lorentz force is carried out by Dirac in 1938 by renormalizing the particle mass. This renormalized equation of motion is called the Abraham-Lorentz-Dirac (ALD) equation of motion. This equation is given by \textcolor[rgb]{0.00,0.07,1.00}{\cite{abraham,lorentz,dirac}}
\begin{equation}
F^\mu_{ALD}=\frac{\mu_0q^2}{6\pi mc}\left[\frac{d^2p^\mu}{d\tau^2}-\frac{p^\mu}{m^2c^2}\left(\frac{dp_\nu}{d\tau}\frac{dp^\nu}{d\tau}\right)\right]\,.
\end{equation}
Equation (21) can now be compared with Eq.(18) of the Jerk equation in curved space. In terms of momentum, $p^\mu$, Eq.(18) can be expressed as
\begin{equation}
F^\mu_A=\frac{\mu_0 q^2}{6\pi mc}\left[\frac{d^2p^\mu}{d\tau^2}+\frac{\Gamma ^\mu_{\,\lambda\nu}}{m}\, p^\lambda\frac{dp^\nu}{d\tau}\,\right]\,.
\end{equation}
Thus, Eq.(22) generalizes the ordinary Abraham - Lorentz force to the case of curved space. It thus  accounts for the electromagnetic radiation emitted by accelerating charge in a gravitational field.

The electromagnetic force on a moving charge $q$ is given by
\begin{equation}
f^\mu=qF^{\mu\nu}V_\nu=m\,a^\mu\,,
\end{equation}
where $F^{\mu\nu}$ is the electromagnetic tensor.
The time rate of change of the acceleration (the Jerk), as a result of the electromagnetic force, is thus
\begin{equation}
J^\mu=\frac{da^\mu}{d\tau}=\frac{q}{m}\,(\partial_\lambda F^{\mu\nu})V_\nu\,V^\lambda+\frac{q^2}{m^2}\,F^{\mu\nu}F_{\nu\lambda}V^\lambda\,.
\end{equation}
The Jerk in a curved space is given by  Eq.(11), which upon using Eq.(23), can be written as
\begin{equation}
J^\mu=\frac{da^\mu}{d\tau}+\frac{q}{m}\, \Gamma^\mu_{\nu\lambda}\,F^{\lambda}_{\,\,\,\alpha}V^\alpha\,V^\nu\,.
\end{equation}
Recall that the Larmor power of an accelerating electric charge  is given by \textcolor[rgb]{0.00,0.07,1.00}{\cite{larmor}}
\begin{equation}
P_L=\frac{\mu_0q^2a^2}{6\pi c}\,,
\end{equation}
while the reaction (Abraham-Lorentz) power is given by
\begin{equation}
P_R=\frac{\mu_0q^2}{6\pi c}\,\vec{v}\cdot\vec{\dot a}\,.
\end{equation}
Since for relativistic case, one has $V^\mu V_\mu=c^2$, then $a^\mu V_\mu=0$ so that $a^\mu\, a_\mu=-\dot a^\mu V_\mu$, or explicitly,
\begin{equation}
 a_\mu a^\mu=-\frac{\dot v^2-(\frac{\vec{v}}{c}\,\cdot\vec{\dot v})^2}{c^4(1-\frac{v^2}{c^2})^3}\,.
\end{equation}
The relativistic power is given by
$$P_L=\frac{2}{3}\,\frac{q^2\gamma^4}{c^3}(a^2_\bot+\gamma^2a^2_\parallel)\,,\qquad |\dot v|^2=a^2_\bot+a^2_\parallel.$$

\section{\textcolor[rgb]{0.00,0.07,1.00}{The gravitational radiation reaction force}}

Owing to the existing analogy between electromagnetism and gravity, one can write the gravitational Larmor power of an accelerating gravitating mass as \textcolor[rgb]{0.00,0.07,1.00}{\cite{gravitomagnetism}}
\begin{equation}
P_G=\frac{2}{3}\, \frac{Gm^2a^2}{c^3}\,,
\end{equation}
where $G$ is the gravitational constant. The gravitational Abraham-Lorentz radiation force will be
\begin{equation}
\vec{F}_G=\frac{2}{3}\, \frac{Gm^2\vec{\dot a}}{c^3}\,\,.
\end{equation}
This can be expressed as
\begin{equation}
\vec{F}_G=\tau_Gm\vec{\dot a}\,\,,\qquad\qquad \tau_G=\frac{2Gm}{3c^3}\,,
\end{equation}
where  $\tau_G$ is some characteristic frequency connected with the emitted gravitational radiation.
For the Earth, one finds $\tau_G=9.83\times 10^{-12}\, s$. This corresponds to a frequency of $\tau_G^{-1}=101\, GHz$. This frequency lies   in the microwave region ($2.95\,\, mm$). Therefore, it may interfere with the cosmic background radiation (CMBR), that is deemed to be a signature of the Big Bang. Thus, a question can be raised whether the CMBR has a gravitational origin rather than electromagnetic? Notice that the characteristic frequency of the electromagnetic radiation emitted by an accelerating charge is exceedingly high and can't be experimentally verified.

It is interesting to note that if $\tau_G$ in Eq.(31) equals to Planck's time, the mass is equal to the Planck's mass ($m_P$). If the Abraham - Lorentz force at Planck's time were equal to the maximal force in the universe, $c^4/4G$ \textcolor[rgb]{0.00,0.07,1.00}{\cite{maximal,maximal1}}, then the Jerk at Planck's time in Eq.(31) would be
\begin{equation}
\dot a_P=\frac{3c^7}{8G^2m_P^2}\,,
\end{equation}
which is  $\dot a\sim 10^{95}\,m/s^3$ that gave birth to our universe!!  However, the value of the Jerk at the present time will be
\begin{equation}
\dot a_0=\frac{3c^7}{8G^2m_0^2}\,,
\end{equation}
where $m_0\sim 10^{53}\, kg$, which amounts to a value of $\dot a_0\sim 10^{26}\, m/s^3$.

At the present epoch of the Universe, Eq.(30) yields
\begin{equation}
\dot a_p=\frac{3c^3}{2R_p^2}\,,
\end{equation}
where $R_p$ is the present radius of the Universe. This amounts to a value of $\dot a\sim 10^{-27} m/s^3$. It is also found that an acceleration of $cH_0$ is prevailing our present space expansion that is hard to explain. This amounts to say that a Jerk of $\dot a=cH^2$ may be connected with such an acceleration. This latter amounts to a value of $\dot a\sim 10^{-27}m/s^3$, where $H\sim 10^{-18}\, sec$. Is the equality of the two Jerks is  a mere coincidence or connected with some deep mystery of  our universe?

\section{\textcolor[rgb]{0.00,0.07,1.00}{Temperature and frequency of the Unruh radiation}}

It was noted by Unruh that an acceleration observer sees a radiation with thermal equilibrium at temperature, $T$, given by \textcolor[rgb]{0.00,0.07,1.00}{\cite{unruh}}
\begin{equation}
T=\frac{\hbar\, a}{2\pi c k_B}\,.
\end{equation}
While in Abraham-Lorentz-Dirac force the acceleration is due to the charge of the particle, and the power of the emitted radiation (Larmor power) is associated with the particle charge too, the radiation due to Unruh is not associated with an accelerating  charged; it is just an observer who is accelerating but still sees a radiation  having a Black-Body nature.  Notice that both effects are independent of the mass of the accelerating particle. Let us assume here that the observer (particle) has a charge that is accelerated by an electric field. Therefore, if the particle acceleration is changing with time, then a reaction force appeared on the charge as shown by Abraham-Lorentz formula \textcolor[rgb]{0.00,0.07,1.00}{\cite{abraham,lorentz}}. Therefore, one would expect the Abraham-Lorentz force to heat the radiation emitted by the accelerating charge.

We could associate here a Larmor power with the radiation emitted by an accelerating electron that radiates according to the Unruh radiation. This will yield a power of
$$\hspace{4cm}P_L=\frac{8\pi^2ke^2}{3c\hbar^2}\, (k_BT)^2\,,\qquad\qquad P_L=\frac{8\pi^2\alpha}{3\hbar}\, (k_BT)^2\,,\hspace{4cm} (i)$$
where $\alpha$ is the fine structure constant.
Applying Eq.(35) in Eq.(17) yields the Abraham-Lorentz force for the Unruh radiation (a thermal force) felt by the accelerating observer as
\begin{equation}
F=\frac{\mu_0k_Bq^2}{3\hbar}\,\,\dot T\,,
\end{equation}
that can be expressed as
\begin{equation}
F=\frac{dp}{dt}\,,\qquad\qquad p=\frac{\mu_0k_Bq^2}{3\hbar}\,\,T\,.
\end{equation}
If we  now connect the above momentum with a de Broglie wavelength of the charge (electron) by, $p=h/\lambda_q$,  then
\begin{equation}
\lambda_q\, T=\frac{6\pi\,\hbar^2}{\mu_0k_Be^2}\,\,\,, \qquad\qquad \lambda_q\, T=\frac{3\hbar\,c}{2\alpha\, k_B}\,\,,\qquad \alpha=\frac{e^2}{2\epsilon_0hc}\,,
\end{equation}
where $\textcolor[rgb]{0.00,0.00,1.00}{\alpha}$ is the fine structure constant. It is interesting that the wavelength of the accelerating electron is connected with the temperature of the radiation it emits. The relation $\lambda_qT=const.$ in Eq.(38) for the matter wave (particle), can be compared with the Wien's displacement for the Black Body radiation, that relates the maximum wavelength of the intensity  distribution to the temperature of the Black Body radiation, by \footnote{$\lambda_{max.}=hc/(xk_BT)$, where $x=4.96511$\,.}
\begin{equation}
\lambda_{max.}\, T=2.89\times 10^{-3}\, \, K m\,\,.
\end{equation}
While the right hand-side of Eq.(39) involves a numeric number, it is a combination of fundamental constants in Eq.(38).
Now Eq.(38) can be written in terms of the frequency, $\nu$, in an analogous manner to that of the Wien's displacement law, as
\begin{equation}
h\nu_q=\frac{4\pi \alpha}{3}\, k_BT\,\,,\qquad\qquad \lambda_q=\frac{6\pi \hbar^2}{\mu_0k_Be^2}\,\frac{1}{T}\,.
\end{equation}
Equation (40) suggests that intensity of matter wave (de Broglie) concomitant with the electron is similar to that of the Black Body radiation. Moreover, one finds
$$\hspace{7cm}\lambda_q=\frac{0.184}{T}\,, \qquad\qquad \hspace{6cm}(A)$$
where $T$ in Kelvin and $\lambda_q$ in meter.
The ratio between the   energy of the matter wave and the thermal energy of the emitted radiation  is
\begin{equation}
\frac{h\nu_q}{k_BT}=\frac{3}{4\pi \alpha}=32.7\,\,\,,
\end{equation}
where $\nu_q=c/\lambda_q$. This indeed shows that the above relation works for high frequency regime. For the Black-Body radiation, the maximum of the intensity occurs such that one has $\frac{h\nu}{k_BT}=x=4.96511$.

Using the Unruh acceleration-temperature equation, Eq.(35), and Eq.(17), one finds
\begin{equation}
\nu_q=\frac{\alpha}{3\pi c}\, a\,\,\,,\qquad\qquad \lambda_q=\frac{3\pi c^2}{\alpha}\,\frac{1}{a}\,.
\end{equation}
Interestingly, we see from Eqs.(35) and (42) that $\lambda_q\propto 1/a\propto 1/T$. Hence, a more accelerating particle emits a cooler radiation than a low accelerating one. Specifically, the frequency (in Hz) of the  matter wave,  as seen by an accelerated observer (charged particle), is  $\nu_q=2.58\times 10^{-8}\,a $, or $\lambda_q=1.162\times 10^{20}/a\,\, (m)$. Equation (42) is very interesting since it couples the  matter wave frequency (de Broglie) of the accelerating charged particle with the temperature of the emitted electromagnetic radiation. Owing to Abraham-Lornetz-Unruh, the kinetic energy of the accelerating electron  is given by
\begin{equation}
E_k=\frac{p^2}{2m}\,\,,\qquad\qquad E_k=\frac{1}{2m}\left(\frac{\mu_0k_Be^2}{3\hbar}\right)^2T^2\,,\qquad\qquad E_k=\frac{1}{2m}\left(\frac{4\pi\alpha k_B}{3c}\right)^2T^2\,,
\end{equation}
upon using Eq.(37). Comparing this with Eq.(i) leads to
$$P_L=\frac{E_k}{\tau_L}\,,\qquad\qquad \tau_L=\frac{\alpha}{3}\,\frac{\hbar}{mc^2}\,.$$
This implies that, if an accelerating electron radiates like Larmor also radiates as Unruh, then this power is equal to the kinetic energy dissipated by the electron during a time given by $\tau_L$ (some uncertainty time). Recall that the ordinary kinetic energy acquired by a particle placed in a heat bath  at temperature $T$ is  $E_k=\frac{3}{2}\,k_BT$, which varies linearly with temperature. The ratio between the rest-mass energy to the thermal energy is
$$\hspace{6cm}\frac{mc^2}{k_BT}=\frac{(4\pi\alpha)^2}{27}\,,\qquad\qquad\hspace{6cm} (B)$$
which is much less than 1 (\emph{viz}., $3.11\times 10^{-4}$). Therefore, the kinetic energy of the accelerating particle, when its rest-mass energy is very much less than its thermal energy,  will be given by Eq.(43). Hence, the electron-positron pairs are freely created by thermal interactions.

If we equate the matter wave frequency in Eq.(42) with the jittery frequency ($2mc^2/\hbar$) of the electron, we set a maximum acceleration that the electron can have, \emph{viz}.
\begin{equation}
a_{max.}=\frac{6\pi}{\alpha}\, \frac{mc^3}{\hbar}\,\,\,,\qquad\qquad a_{max.}=24\pi^2\,\frac{mc^2}{\mu_0e^2}\,.
\end{equation}
This can be associated with a maximum electric field  by which a charge (electron) is accelerated, $a=(e/m)\, E$, \emph{i.e}.,
\begin{equation}
E_{max.}=\frac{6\pi}{\alpha}\,\frac{m^2c^3}{e\hbar}\,,\qquad\qquad E_{max.}=24\pi^2\,\frac{m^2c^2}{\mu_0e^3}\,.
\end{equation}
It is now interesting to compare the above maximum electric field  with the Schwinger  electric field limit, $E_0=m^2c^3/e\hbar$,  beyond which the electromagnetic field is thought to become nonlinear \textcolor[rgb]{0.00,0.07,1.00}{\cite{schwinger,schwinger2}}. The electric field in Eq.(45) exceeds the Schwinger limit by three orders of magnitude \textcolor[rgb]{0.00,0.07,1.00}{\cite{schwinger1}}. The maximum acceleration in Eq.(44) amounts to $a_{max.}=6.01\times 10^{32}\, m/s^2$\,.

Let us now apply Eq.(44) in Eq.(35) to obtain the maximum temperature limit
$$T_{max.}=\frac{3}{\alpha}\, \frac{mc^2}{k_B}\,.$$
For an accelerating electron, one finds $T_{max.}=2.44\times 10^{12}\, K$. This temperature amounts to a value of $0.21\, GeV$. Interestingly, a Planck temperature of $10^{32}K$ could be radiated by an accelerating charged particle with a mass of $\sim 10^{-9}kg$. Using Eqs.(44) and (42), the minimum de Broglie wavelength of the the accelerating electron is $\lambda_{min.}=\frac{\hbar}{2mc}$.

Substituting Eq.(43) in Eq.(26) yields the corresponding maximum Larmor power will be
\begin{equation}
P^{max.}_L=96\pi^3\,\frac{m^2c^3}{\mu_0e^2}\,\,.
\end{equation}
However, a maximal gravitational power and force in the universe are respectively  given by
\begin{equation}
P_{max.}=\frac{c^5}{4G}\,\,,\qquad\qquad F_{max.}=\frac{c^4}{4G}\,.
\end{equation}
Using Eq.(35), the black body intensity of the radiation emitted by an accelerating observer will be
\begin{equation}
I=\frac{\hbar\,\,a^4}{960\,\pi^2c^6}\,.
\end{equation}

\subsection{\textcolor[rgb]{0.00,0.07,1.00}{Hawking radiation}}

In 1974 Hawking  derived a formula that relates the black hole radiation to its mass. It is a quantum relation as that of Unruh. It states that black holes radiate like a black body radiation with temperature, T. It is given by \textcolor[rgb]{0.00,0.07,1.00}{\cite{hawking}}
\begin{equation}
k_BT=\frac{\hbar c^3}{8\pi GM}\,,
\end{equation}
where $M$ is the mass of the Black hole. Intrigued by Eq.(35), Eq.(48) can be casted in the form
\begin{equation}
k_BT=\frac{\hbar a_B}{2\pi c}\,,\qquad\qquad a_B=\frac{c^4}{4 G M}\,,
\end{equation}
where $a_B=\kappa$, is known as the surface gravity. Following the same line of reasoning that gave rise to Eq.(30), one finds the radiation force (Abraham-Lorentz force) for the Black Hole to be determined by
\begin{equation}
F_B=-\frac{c}{6}\, \dot M\,,
\end{equation}
which is independent of $G$, implying that this force is a pure relativistic effect. The above force could be the reason behind the fact that a black hole can't evaporate completely, since as the mass decreases an increasing positive force is experienced, while mass increase results in a negative force ceasing the emitted radiation. One can associate a power with the Abraham-Lorentz force that is $P_A=-\dot M c^2/6$ which is 1/6 of the rest-mass power.

It is interesting the reaction for depends only on the rate of evaporation of the black hole ($\dot M$). Using the energy-mass equivalent equation, the power will be expressed by $P=-dE/dt=\dot Mc^2$. Hence, the Abraham-Lorentz reaction force is proportional to the power delivered by the black hole due to its radiation.

The entropy ($S$) of the black hole is related to its horizon area ($A)$ by the Bekenstein's  formula \textcolor[rgb]{0.00,0.07,1.00}{\cite{entropy}}
\begin{equation}
S=\frac{k_Bc^3A}{4G\hbar}\,\,.
\end{equation}
The horizon area of the black hole is given by $A=16\pi G^2M^2/c^4$, so that Eqs.(30), (35) and (50) yield
\begin{equation}
S=\frac{\hbar c^5}{16 \pi G k_B}\frac{1}{T^2}\,.
\end{equation}
Upon using Eq.(49), Eq.(52) yields
\begin{equation}
 S=k_B\left(\frac{4\pi GM^2}{\hbar c}\right)\,.
\end{equation}
Recall that if we define the angular momentum (spin) of the black hole by \footnote{The spin for a spherical mass is defined by $J=I\omega$, where the moment of inertia $I=2/5 MR^2$,  and for a black hole $\omega=c/R_s$ and $R_s=2GM/c^2$, so that $J=4GM^2/5c$.}
$$J=\frac{4\pi GM^2}{c}\,,$$
so that Eq.(53) can be written as
\begin{equation}
 S=\frac{k_B}{\hbar}\, J\,.
\end{equation}
It is remarkable to see from Eq.(54) that the entropy of the black hole is a measure of its  spin angular momentum. Now one can combine Eqs.(49) and (53) to write
\begin{equation}
 S\,a_B=\pi k_B\left(\frac{Mc^3}{\hbar}\right)\,.
\end{equation}
Upon using Eqs.(50) and (53), the entropy can be associate with the back reaction force of an accelerating particle as
\begin{equation}
F_B=\frac{\hbar c^2}{48 \pi k_B GM}\, \dot S\,,
\end{equation}
which can advocate a momentum of the form
\begin{equation}
p_B=\frac{\hbar c^2}{48 \pi k_B GM}\,  S\,,
\end{equation}
Note that while the entropy density of the black body radiation is $s=\frac{4\pi^2k_B^4}{45c^3\hbar}\, T^3$, it is, $s=16 \pi \frac{k_B^2c^2}{\hbar^2G}\, T$ for black hole radiation suggesting that it is a gravitational radiation. This shows that black hole and black body radiations are dissimilar. Since as the black hole radiates its mass decreases and therforee its temperature increases, it should reach a maximum temperature that can be obtained by equating $T_{max.}$ with that in Eq.(48). This yields a left-over mass (minimum mass)  of
\begin{equation}
M_f=\sqrt{\frac{\alpha}{24\pi }}\left(\frac{c\hbar}{G}\right)^{1/2}\,.
\end{equation}
It is thus very remarkable that black holes don't evaporate completely. The remnant part of black hole is to preserve mass and spin. The final spin of the black hole is $J_{min.}=(\alpha/6)\hbar$, and upon Eq.(54) one finds the minim entropy to be, $S_{min.}=(\alpha/6)k_B.$ The final radius of the black hole will be $R_f=\sqrt{\frac{\alpha G\hbar}{6\pi c^3}}$. These minimal quantities would be the precursor (seed) of the future quantum fluctuations that can build observable effects.
Interestingly, the maximum acceleration in Eqs.(44) and (47) are approximately equal for a mass given by
\begin{equation}
m_c=\sqrt{\frac{\alpha}{24\pi}}\,\left(\frac{\hbar c}{G}\right)^{1/2}\,,
\end{equation}
which can be compared with the Planck's mass ($M_p=\sqrt{\frac{\hbar c}{8\pi G}}$). It can be seen as the mass of a charged particle at which the gravitational force is equal to the electric force. It is surprising  that the two masses in Eq.(58) and (59) are equal.

Applying Eq.(35) in Eq.(26) yields
\begin{equation}
P_L=\frac{2\pi e^2}{3\varepsilon_0c\hbar^2}\,(k_BT)^2\,,
\end{equation}
which when compared with Eq.(43) yields
\begin{equation}
P_L=\frac{E_k}{(\tau_q/2)}\,.
\end{equation}
Thus, the power $P_L$ can be seen as the kinetic energy dissipated during a time $\tau_q/2$. Now apply Eq.(35) in Eq.(29) and (43)  to obtain
\begin{equation}
P_G=2\tau_G\left(\frac{c}{r_e}\right)^2\,E_k\,,\qquad\qquad P_G=\frac{E_k}{\tau_{qG}}\,,\qquad \tau_{qG}=\frac{9}{8}\frac{\tau_q^2}{\tau_G}\,.
\end{equation}
It seems that the time $\tau_{qG}$ is a competing time between charge and mass impact on the  emitted radiation, \emph{i.e}., a competition between electromagnetic and gravitational waves.

\subsection{\textcolor[rgb]{0.00,0.07,1.00}{Cosmological radiation}}

If we assume the black body radiation relation, $TR=const.$, where $R$ is the radius of the universe (scale factor),  still holds, today, then $T_0R_0=T_pR_p$, where the subscripts ``0" and ``$p$" stand for Planck and today values. This implies that $T_0\sim 10^{-29}K$, where $R_p\sim 10^{-35}m\,, T_p\sim 10^{32}K$. If the object emitting that radiation follows the Unruh radiation, Eq.(35), then one finds $T\sim 10^{-29}K$ if it accelerates today at $a\sim 10^{-10}m/s^2$. This  amounts to an energy of $10^{-33}eV$ that is attributed to the mass of the graviton \textcolor[rgb]{0.00,0.07,1.00}{\cite{mass1,mass2}}. This latter acceleration is found to be related to the cosmic expansion of the space permeating the whole universe. It is thus interesting that it is the space that accelerates, and thus produces a black body radiation permeating the whole space having a temperature of $\sim 10^{-29}K  \,(10^{-33}eV)$, today \textcolor[rgb]{0.00,0.07,1.00}{\cite{quantum_arbab,caroll}}.  Such a background energy is recently attributed to the graviton mass, or a cosmological constant. It could also account for the mass of the gravitational wave \textcolor[rgb]{0.00,0.07,1.00}{\cite{graviton}}.

\section{\textcolor[rgb]{0.00,0.07,1.00}{Concluding remarks}}

It was shown by Abraham-Lorentz-Dirac that an accelerating charged particle experiences a reaction force (radiation force) proportional to the particle's Jerk (time rate of change of the acceleration). We explored in this the paper the extension of the Jerk when the particle is accelerating in a curved space. As regards to Abraham-Lorentz-Dirac force, we studied the de Broglie wave associated with the accelerating charge. We further assume that the Unruh radiation due to accelerating observer (charge) causes a reaction force on the observer as that of the Abaraham-Lorentz force. The intensity of the matter wave is found to follow a law that is  analogous to  the Wien's displacement law. Equating the matter wave frequency to the jittery frequency, advocated by Schrodinger as exhibited by an electron motion, we obtained a maximal acceleration that limits the particle acceleration. A maximal  Larmor power of the radiation emitted is also derived. A maximal electric field accelerating the charged is set that is found to exceed the Schwinger electric field limit. A black hole emitting Hawking radiation will end in a final state where its mass, entropy and temperature are specified. The universe today is shown to be permeated by a black body Unruh radiation of temperature $10^{-29}K (10^{-33}eV)$, owing to its space acceleration.

\end{document}